\documentstyle[prd,aps]{revtex}
\begin{document}
\input epsf \renewcommand{\topfraction}{0.8}


\def\d{{\rm d}}
\def\be{\begin{equation}}
\def\ee{\end{equation}}
\def\lcdm{\Lambda{\rm CDM}}
\def\pdm{\phi{\rm DM}}
\def\mp{m_{\rm p}}
\def\op{\Omega_{\phi}}
\def\om{\Omega_{\rm m}}
\def\ob{\Omega_{\rm b}}
\def\odm{\Omega_{\rm DM}}
\def\ocdm{\Omega_{\rm CDM}}
\def\ol{\Omega_{\Lambda}}
\def\oq{\Omega_{\rm Q}}
\def\ot{\Omega_{\rm tot}}
\def\etal{et al.~}
\def\prd{Phys. Rev. D~}
\def\rhob{\rho_{\rm b}}
\def\rhom{\rho_{\rm m}} 
\def\rhop{\rho_{\phi}}
\def\opp{\omega_{\phi}}
\def\pp{p_{\phi}}
\catcode`@=11
\def\gsim{\mathrel{\mathpalette\@versim>}}
\def\@versim#1#2{\lower0.2ex\vbox{\baselineskip\z@skip\lineskip\z@skip
      \lineskiplimit\z@\ialign{$\m@th#1\hfil##$\crcr#2\crcr\sim\crcr}}}
\catcode`@=12
\catcode`@=11
\def\lsim{\mathrel{\mathpalette\@versim<}}
\def\@versim#1#2{\lower0.2ex\vbox{\baselineskip\z@skip\lineskip\z@skip
       \lineskiplimit\z@\ialign{$\m@th#1\hfil##$\crcr#2\crcr\sim\crcr}}}
\catcode`@=12
\draft
\twocolumn[\hsize\textwidth\columnwidth\hsize\csname
@twocolumnfalse\endcsname


\title{Dark energy and dark matter from an inhomogeneous dilaton}

\author{Mikel Susperregi}

\address{School of Mathematical Sciences, Queen Mary, 
University of London, London E1 4NS, UK \\ 
{\rm m.susperregi@qmul.ac.uk}}

\date{\today}
\maketitle
\begin{abstract}
{A cosmological scenario is proposed where the dark matter (DM) 
and dark energy (DE) of the universe are two simultaneous  
manifestations of an inhomogeneous dilaton. The equation of state 
of the field is scale dependent and pressureless at galactic 
and larger scales and it has negative pressure as a DE at very large scales. 
The dilaton drives an initial inflationary phase 
followed by a kinetic energy-dominated 
one, as in the ``quintessential inflation'' model introduced 
by Peebles \& Vilenkin, and soon after the end of inflation 
particle production seeds the first inhomogeneities 
that lead to galaxy formation. The dilaton is trapped near the 
minimum of the potential where it oscillates like a massive 
field, and the excess of kinetic energy is dissipated 
via the mechanism of ``gravitational 
cooling'' first introduced by Seidel \& Suen.  
The inhomogeneities therefore behave like solitonic 
oscillations around the minimum of the potential, known 
as ``oscillatons'', that we propose account for most DM in galaxies. 
Those regions 
where the dilaton does not transform enough kinetic 
energy into reheating or carry an excess of it 
from regions that have cooled, evolve to the tail of the 
potential as DE, driving the acceleration of the universe.} 
\end{abstract}

\vskip2pc]

\section{Introduction}

We live in a universe where there is an abundance of DM 
and DE that we do not see or understand. Fortunately, 
progress in cosmology has been swift in determining parameters 
quantitatively, and CMB measurements give us the most accurate 
ratios of the cosmic recipe. However, qualitatively 
we have an enormous vacuum. This paper proposes a model to close 
this gap, though far from explaining cosmology any better than 
other models, it has the virtue of simplicity. 

The model consists of one scalar field, the dilaton. $\phi$ 
drives an initial inflationary phase, followed by a kinetic energy 
(KE) dominated expansion, in an evolution similar to the
``quintessential inflation'' 
model of Peebles \& Vilenkin \cite{peebles2}. Around the end of 
inflation (EOI), the transition results in gravitational particle 
production, a phenomenon first described by Ford \cite{ford} and Spokoiny 
\cite{spokoiny} (also used in \cite{peebles2}). The dilaton releases
gravitational energy to produce matter as well as entropy
fluctuations. We propose that this process is unevenly efficient and it 
introduces inhomogeneities. A further and more significant source of 
inhomogeneities is introduced by a feature in the potential, which 
is in the form of a ``trough'' prior to the decaying 
tail of the potential. The decaying field rolls down to the minimum 
of the ``trough'', 
where it dissipates KE in some measure by reheating 
and more importantly by ``gravitational cooling'', 
a process first discovered by Seidel \& Suen \cite{seidel2} 
in the context of boson stars. 
This process is key to our model so that the dilaton can get rid of 
enough KE to remain consigned to the ``trough'' of the potential, 
where it oscillates like a massive field. Seidel \& Suen call this 
oscillating field an ``oscillaton''. These inhomogeneities account 
for the DM content, and the dilaton here behaves like a soliton, i.e. 
a massive scalar that is confined to a finite region of space, is
non-singular and non-topological. 
On the other hand, regions where the KE does not 
dissipate quickly enough either by the reheating or ``gravitational 
cooling'' move on to the tail of the potential and to the 
DE-dominated era. Here $\phi$ behaves as in ``extended quintessence'', 
first discovered by Perrotta, Baccigalpi \& Matarrese 
\cite{perrotta3}. These models of DE with scalar-tensor 
gravity have been widely studied in the literature 
(see
\cite{amendola2,baccigalupi2,bartolo,chiba,chiba2,esposito,perrotta2,riazuelo,ritis,uzan}). 

In a nutshell then, our proposal is a universe that consists 
of baryons and scalar-tensor gravity. It is certainly a 
concession to credibility that both DM and DE   
are caused by $\phi$, for which there exists no laboratory 
or direct evidence. There is a case to be made however that 
a universe with fewer parameters and the least number of 
fields is preferred over more complicated models. Recently 
a model with one tachyonic field explaining both DM and DE 
was developed by Padmanabhan \cite{padmanabhan} 
(also \cite{bagla,padmanabhan2}). Also, there are various motivations  
for using the ``extended'' model: the non-minimal coupling 
$R\phi$ of the dilaton does arise in the quantization of fields in curved 
spacetime \cite{ford,birrell}, and in multidimensional theories 
\cite{maeda}, such as superstrings and induced gravity. 

CMB and LSS observations give us a 
measurement of DM and DE that is roughly in a ratio 
of 3/7.\footnote{The first-year WMAP data analysis 
\cite{WMAP} concludes that $\ot=1.02\pm 0.02$, DE  
$\oq=0.73\pm 0.04$, matter content $\om=0.27\pm 0.04$, of which 
$\ob=0.044\pm0.004$. I.e. 4.4\% baryons, 22\% DM  
and 73\% DE. On the other hand, the DASI instrument 
analysis \cite{DASI} yields 5\% baryons, 35\% DM and 
60\% DE. We adopt 5\% baryons, 
25\% DM and 70\% DE \cite{lee,stompor,bernardis,hu3}.}
We use this not so much to choose a potential but to test 
its plausibility, bearing in 
mind that 30\% of the energy of the dilaton is ``trapped'' in an 
``oscillaton'', so it manifests itself as DM, and 
the remaining 70\% moves on to the tail of the potential to trigger 
cosmic acceleration as DE.

\section{Theory}

\subsection{Dilaton potential}

The dilaton has the following potential  
\[
V(\phi<0)= \lambda(\phi^4+M^4), 
\]
\be \label{V}
V(\phi\geq 0) = \frac{\lambda M^{12}}{\phi^8 +M^8}
\Big\{\beta \Big[\Big(\frac{\phi}{\phi_0}\Big)^2-1\Big]^2
+(1-\beta)\Big\},
\ee
where $\lambda$ is determined by observation and $M$, 
$\phi_0$ and $\beta$ by the theory. LSS estimates give 
$\lambda\approx 10^{-14}$ \cite{linde}. While 
$\phi\lsim -\mp$ the dilaton triggers chaotic 
inflation $V\sim\lambda\phi^4$ 
\cite{peebles2,linde}; for $\phi\gsim\mp$, it behaves as 
quintessence 
\cite{peebles2,perrotta3,riazuelo,carroll,coble,steinhardt,ratra,wetterich,zlatev}.
The $\phi<0$ branch of (\ref{V}) is identical 
to the ``quintessential inflation'' 
model of Peebles \& Vilenkin \cite{peebles2}, and the $\phi\geq 0$ 
branch retains its asymptotic behaviour (for $\beta>0$)
\footnote{c.f. $V(\phi\geq 0)=\lambda M^4/(\phi^4+M^4)$ in 
\cite{peebles2}.}
\be
V(\phi\to\infty)\sim\Big(\frac{\lambda M^{12}\beta}{\phi_0^4}\Big)
\phi^{-4},
\ee
and at intermediate values $\phi\sim M$, the 
potential has a ``trough'' with a minimum at $\phi=\phi_0$.\footnote{For 
$\beta=1$; in the case $0<\beta<1$, the minimum is slightly 
shifted to the right.} The value 
of $\beta$ determines the depth of the trough ($0\leq\beta\leq 1$). 
The two extremes are $\beta=1$, where we have a ``deep trough''      
and $\phi=\phi_0$ is an absolute minimum, and $\beta=0$ which 
is a ``no-trough'' potential (there is one saddle point, no minima) 
which tails off $\sim\phi^{-8}$.  

The ratio $\nu\equiv\phi_0/M$ controls the location of the trough and the 
height of the hill ($0<\nu< 1$; if $\nu\gsim 1$ the potential 
does not have a trough feature). 
On the other hand $\beta$ has a smoothing effect 
over both the trough and the hill. A ``deep trough'' is followed 
by a ``high hill'', and decreasing $\beta$ results in decreasing 
both the depth of the trough and the height of the hill. 
In Fig.1 we have plotted $\nu =0.64$, a case 
where the height of the hill is $\sim \lambda M^4$. 
The shape of the curve is chiefly sensitive to variations of $\nu$. 
For instance, $\nu =0.60$ doubles the height of 
the hill, and $\nu =0.80$ reduces it by $2/3$. The 
top of the hill is located at $\phi\sim M$, and the amplitude is 
\be 
V_{\rm hill}\approx \frac{\lambda M^4}{2}
\Big(\frac{1}{\nu^2}-1\Big)^2.
\ee
The parameters $\nu$ and (to a lesser extent) $\beta$ 
have to be therefore just right to yield the right ratio of DE  
to DM. If $\nu$ is just a bit too small (and not aided by 
a small $\beta$ to diminish the features), then the dilaton everywhere  
is trapped as an oscillaton in the trough and it manifests itself purely as 
DM. On the other hand, if $\nu$ is too big then the dilaton 
simply rolls down to end up entirely as DE and 
we obtain no DM (other than that produced by gravitational 
particle production at EOI). Roughly we need a deep trough 
that retains 30\% of the energy of the field and allows the 
remainder to progress down the tail of the potential 
to dominate as DE. 

As we shall see in the following sections, 
the theory predicts both $M$ and $\nu$. 

\begin{figure}[t]
\centering
\leavevmode\epsfysize=6.3cm \epsfbox{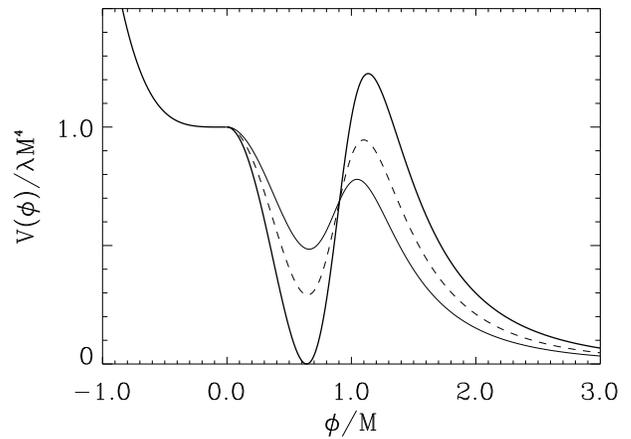}\\
\vskip 0.2cm
\caption[fig1]{$V(\phi)$ for $\beta=0.5,0.7,1.0$. 
The case $\beta=1.0$ is a ``deep trough'' ({\it thick solid 
line}), $\beta=0.5$ a ``shallow trough'' ({\it thin solid line}), 
and $\beta=0.7$ is an intermediate trough 
({\it dashed line}). Here $\nu =0.64$.}
\end{figure}

\subsection{Dynamics}

The dilaton is described by the action 

\be \label{action}
{\cal S}= \int\!\d^4x\,\sqrt{-g}\Big[{\mp^2\over 16\pi}R
-{1\over 2}\xi\phi^2 R-{1\over 2}\phi_{;\mu}\phi^{;\mu} 
-V\Big],
\ee
where $\mp$ is the Planck mass, $R$ is the Ricci scalar, 
$\xi$ is the coupling of the field (assumed universally constant),  
and $V(\phi)$ the dilaton potential   
described in \S II.A. 
The equations that derive from (\ref{action}) 
have been widely studied in the 
literature  
\cite{perrotta3,amendola2,baccigalupi2,bartolo,chiba,chiba2,esposito,perrotta2,riazuelo,ritis,uzan,abreu,accetta,amendola,damour2,hwang,nagata},  
and here we will use the essentials that are relevant to our 
study. 

The field equation is 
\be \label{KG}
\Box\phi-\xi R\phi-V^\prime =0,
\ee
where the prime denotes $d/d\phi$. The Einstein's equations are 
\be 
R_{\mu\nu}-\frac{1}{2}g_{\mu\nu}R=\kappa_{\rm eff} T_{\mu\nu},
\ee
where $\kappa\equiv 8\pi G$ and the effective gravitational 
constant $\kappa_{\rm eff}=\kappa (1- \kappa \xi\phi^2)^{-1}$ 
(following \cite{abreu}), 
or analogously the effective Planck mass\footnote{For 
$\xi>0$ there is a potential anomaly  
$m_{\rm p,eff}^2\to 0$ in the limit $|\phi|\to\infty$, see Fig.2. 
To have a viable cosmology  $m_{\rm p,eff}\neq 0$, 
we need $\phi\ll \phi_{\rm crit}=\mp/\sqrt{3\xi}$ (the 
smallness of $\xi$ is confirmed by observations, 
$|\xi|\lsim {\cal O}(10^{-2})$ following  
\cite{perrotta3,baccigalupi2,chiba,gillies});
non-negligible departures from $\mp$ will take place during the  
inflationary and DE-dominated phases, when 
$\phi^2\gsim \mp^2$.}
\be \label{planckmass}
m_{\rm p,eff}^2=\mp^2 (1-3\xi\phi^2/\mp^2).
\ee
It is $m_{\rm p,eff}$ that is measured experimentally. 
At present, $m_{\rm p,eff}=4.2\times 10^{18}$ GeV. 
The energy momentum tensor is  
\[
T_{\mu\nu}=\phi_{;\mu}\phi_{;\nu}
-{1\over 2}g_{\mu\nu}(\phi_{;\rho}\phi^{;\rho}
+2V)
\]
\be
+\xi\Big[g_{\mu\nu}\Box(\phi^2)-(\phi^2)_{;\mu ;\nu}\Big].
\ee
For the spatially homogeneous component of $\phi$, (\ref{KG}) becomes 
\be \label{KGhom}
\ddot\phi+3H\dot\phi +6\xi (\dot H+2H^2)\phi +V^\prime =0,
\ee 
where 
\be
H^2=m_{\rm p,eff}^{-2}(\rhop+\rhom),
\ee
\be 
R=6(\dot H+2H^2),
\ee
and 
\be \label{rho}
\rho_{\phi}=\frac{1}{2}\dot\phi^2+V
+ 3H\xi \phi (H\phi + 2\dot\phi),
\ee
\[ 
p_{\phi}=\frac{1}{2}\dot\phi^2-V
-\xi\Big[(2\dot H+3H^2)\phi^2 
\]
\be \label{pressure}
+4H\phi\dot\phi+2\phi\ddot\phi+2\dot\phi^2\Big],
\ee
where we have used $G_{00}=\kappa_{\rm eff}a^2\rhop$ and 
$\pp=\frac{1}{3}T_{ij}\delta^{ij}a^{-2}$ (see \cite{uzan}). 
The conservation of the field $T^{\mu\nu}_{;\mu}=0$ reduces to 
\be 
\dot\rhop+3H(\rhop+\pp)=0,
\ee
which is equivalent to (\ref{KGhom}). The equation of state 
is given by 
\be \label{state}
\opp= \frac{\pp}{\rhop},
\ee 
which is a scale-dependent quantity in our scenario, once 
inhomogeneities set in. 

\begin{figure}[t]
\centering
\leavevmode\epsfysize=6.3cm \epsfbox{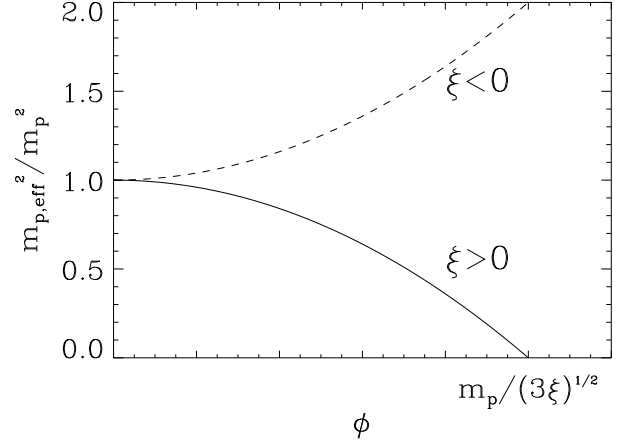}\\
\vskip 0.2cm
\caption[fig2]{$m_{\rm p,eff}$ is a slowly-varying 
monotonically decreasing (increasing) function for 
$\xi>0$ ($\xi<0$).}
\end{figure}

\section{KE domination}

The $V(\phi<0)$ branch in (\ref{V}) is identical to  
that of \cite{peebles2}, so our solutions at the 
onset of KE domination can only differ in the effect 
caused by $\xi\neq 0$. Let us quantify this difference. 

Soon after EOI, which is at $\phi\approx -\mp$ (the slow-roll parameters 
are $\epsilon\approx 8\mp^2\phi^{-2}$ and 
$\eta\approx 1.5 \epsilon$), most of the potential energy 
is converted into KE leading to a ``deflationary'' phase, 
following Spokoiny \cite{spokoiny} 
(also named as ``kination'' by Joyce \cite{joyce}). During this 
phase $\rhop\sim \dot\phi^2/2$. Therefore the solution 
$\dot\phi\propto a^n$ fully determines the energy 
density. It is 
\be \label{n}
n= -3\frac{1+2\xi\psi}{1+3\xi\psi}
\ee 
where 
\be \label{psi}
\psi =2\big(6\xi+ \sqrt{36\xi^2-12\xi+2}\big)/(1-6\xi)
\ee
for $\xi\neq 1/6$, and $\psi(1/6)=-1$. 
For $\xi\ll 1/2\sqrt{2}$, the solution is 
\be 
n\approx -3+6\sqrt{2}\xi.
\ee
Therefore 
\be \label{KE}
\rhop\approx \lambda \mp^4 \alpha (a/a_*)^{2n},
\ee 
where $a_*$ is the expansion factor at EOI and 
\be 
\alpha=(1-3\xi)(1-6\xi-12\xi\psi^{-1})^{-1}.
\ee 
For small $\xi$, 
\be 
\rhop\approx \lambda \mp^4 (1+7.24\xi)
(a/a_*)^{-6+16.97\xi},
\ee
($\xi=0$ recovers the result of \cite{peebles2}  
$\rhop\approx \lambda\mp^4 (a/a_*)^{-6}$). Also 
\be \label{field}
\phi\approx \mp\Big[\sqrt{2\alpha} 
\,{\rm ln}\Big(\frac{a}{a_*}\Big)-1\Big],
\ee
and the expansion factor $a\sim t^{-1/n}$. The potential 
during this period varies as 
\be
V\approx \frac{3\lambda M^8}{4\alpha^2}\,\mp^{-4}
\Big[\ln \Big(\frac{a}{a_*}\Big)\Big]^{-4}.
\ee

\begin{figure}[t]
\centering
\leavevmode\epsfysize=6.2cm \epsfbox{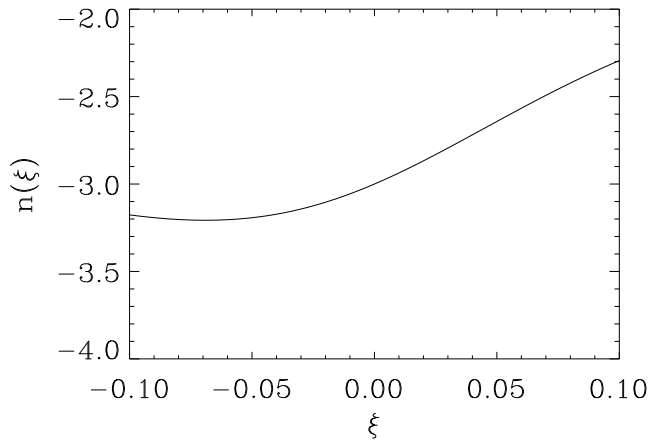}\\
\vskip 0.2cm
\caption[fig3]{The solution
(\ref{n})(\ref{psi}) is slowly varying and correctly 
reproduces the result $n(0)= -3$.}
\end{figure}

For realistic values of $\xi$ derived from observational constraints 
on $|\dot G/G|$ (as used in \cite{perrotta3,baccigalupi2,chiba,gillies}
and references therein), $|\xi|\lsim 1.0-2.2\times 10^{-2}$, this 
has a fairly small impact in $n(\xi)$ as one can see from Fig.3. 
For $\xi$ of a few percent, this places $n$ in the range 
$-3.2\lsim n\lsim -2.8$. Realistically the most significant 
change results in $n\approx -2.7$, therefore $\rhop\sim a^{-5.4}$, 
which is not a very dramatic difference with respect to 
$\sim a^{-6}$ and even this entails pushing the 
constraints on $\xi$ to about $4.2\times 10^{-2}$ (within the 
bounds of \cite{perrotta3,baccigalupi2,chiba,gillies}, 
the solution lies between the curves 
$(a_*/a)^{6.4}\lsim \rhop/\lambda\mp^4 \lsim (a_*/a)^{5.6}$). 
Therefore the energy density 
of the field during this stage decreases faster than that of radiation,
and this point, made by Ford, Spokoiny and Peebles \& Vilenkin 
\cite{ford,spokoiny,peebles2}, stands here too. 

A more marginal point is that the amplitude of $\rhop$ is 
only mildly dependent on $\xi$ (it varies up to  
$0.8-1.3$ times its $\xi=0$ value), and (\ref{field}) 
can vary up to about 20\% with respect to $\xi=0$. 

The pressure of the dilaton is 
\be
\pp\approx \lambda \mp^4\gamma (a/a_*)^{2n},
\ee
where 
\be
\gamma =(1-4\xi)-2\psi(2+n)\xi
-\psi^2\Big(\frac{3}{2}+n\Big)\xi.
\ee
Therefore (\ref{state}) becomes
\be \label{opp}
\opp=\gamma\alpha^{-1}.
\ee
These results are shown in Fig.4. Both $\alpha$ and 
$\gamma$ are of order unity for realistic values 
of $\xi$, as is $\opp$ (e.g. $\opp(0.03)\approx 1.25$).  

\begin{figure}[t]
\centering
\leavevmode\epsfysize=6.2cm \epsfbox{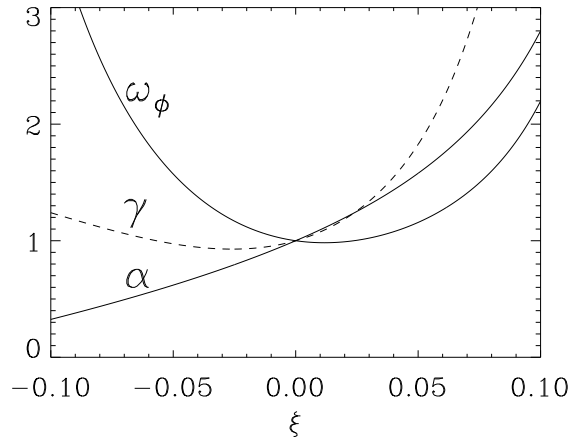}\\
\vskip 0.2cm
\caption[fig4]{$\opp$ as a function of $\xi$ ({\it thick solid line}). 
$\alpha$ ({\it thin solid line}) and $\gamma$ 
({\it dashed line}) 
are the coefficients of the field energy density and pressure 
respectively. The departure from unity (at $\xi=0$) for all 
three functions is slight for realistic values of $\xi$.}
\end{figure}

\section{Particle production}

As pointed out by Ford \cite{ford}, there is particle 
production owing to the transition of the metric 
at EOI. The relevant QFT to calculate the $\rhom$ created 
can be found in \cite{ford,birrell}. 
It has been worked out by Ford, Spokoiny and Peebles \& Vilenkin  
\cite{peebles2,ford,spokoiny}, that gravitational particle production 
is capable of reheating the universe after inflation and being 
the dominant source of matter.  

In our scenario, we do not require the mechanism to create {\it all} 
matter, as most of it is in the form of the 
oscillaton. We propose 
that particle production is inhomogeneously 
efficient, and the inhomogeneities are caused by a stochastic 
parameter $\eta\geq 0$. 
For $\eta=0$ no particles are produced, and for $\eta=1$ the 
energy density of the created particles is given as in 
\cite{peebles2,ford,spokoiny}. We also allow for the possibility 
$\eta > 1$. Following Ford \cite{ford} and others 
\cite{peebles2,damour4,giovannini}, 
\be
\rhom\sim \lambda^2 \mp^4 \eta\,\alpha^2  R\Big(\frac{a_*}{a}\Big)^4,
\ee 
for the created particles and $R\sim 10^{-2}$. We consider 
the number of scalars $N_S=1$ and arbitrary (though small) $\xi$. 
The energy of the created particles is as in Spokoiny \cite{spokoiny} 
$\epsilon\sim H_*(a_*/a)\sim 10^{12}~{\rm GeV}$, their 
number density $n\sim R\epsilon^3$, and thermal 
equilibrium is achieved at about 
\be 
\Big(\frac{a_{\rm th}}{a_*}\Big)\sim 10^2-10^3,
\ee
at a temperature 
\be
T_{\rm th}\sim \rhom^{1/4}(a_{\rm th})\sim (10^9-10^{10})
\eta^{1/4}~~{\rm GeV}.
\ee

The exact local amplitude of $\eta$ is not important, but 
we consider two types of regions, where $\eta$ 
takes the values $\eta_-$ and $\eta_+$ (small and large), 
and the scale of these regions is $D$. In the $D_+$ regions 
there is enough particle production to trigger further 
reheating, and finally most of the KE is expelled by 
means of ``gravitational cooling'', as we will 
describe in \S VII. The dilaton in these regions 
is then trapped in the trough, 
oscillating around $\phi\approx \phi_0$. The $D_-$ 
regions do not dissipate enough KE and they 
progress to the tail of the potential until they finally 
become DE-dominated. 

\subsection{Radiation era}

The onset of the radiation era is given by $\rhom\sim\rhop$, 
and given the inhomogeneities created by $\eta$, this varies 
spatially (leaving also the possibility open for skipping the 
radiation era altogether in sufficiently large regions 
where $\eta\approx 0$). The ratio of the energy density 
of the created particles and that of the dilaton 
is 
\be
\frac{\rhom}{\rhop}\sim \lambda R \eta
\Big(\frac{a_*}{a}\Big)^{4+2n},
\ee
(the result of \cite{peebles2} is recovered  
$4+2n(0)=-2$, $\eta=1$). Therefore 
\be \label{ara*}
\frac{a_r}{a_*}\sim (\lambda R\eta)^{1/(4+2n)}\sim 10^8\eta^{-1/2},
\ee  
and a temperature $T_r\sim 10^3 \eta^{1/2}~{\rm GeV}$. 
We require that $a_r$ precedes the DE-dominated phase (if 
radiation domination is to take place at all), and 
we will examine this condition in the next section.

\section{DE domination}

The potential becomes dominant at a scale factor
\be  \label{a-phi}
\frac{a_\phi}{a_*}\sim \Big(\frac{\phi_0^4 \mp^8}{M^{12}}\Big)^{-1/2n}
{\rm ln}^{-2/n}\Big(\frac{\phi_0 \mp^2}{M^3}\Big),
\ee
at which point $D_-$ regions enter a period of accelerated 
expansion. As was shown by \cite{peebles2}, the value of the 
field during this phase remains roughly constant, and it 
is approximately the value $\phi_r$ at the time $a\approx a_r$. 
This is 
\be
\phi_r\approx \sqrt{2\alpha}\,\mp \frac{1}{4+2n}\,\ln(\lambda R\eta),
\ee
and at present
\be
\rhop\approx V(\phi_r)
\approx \lambda \beta \frac{M^{12}}{(\phi_0\phi_r)^4}
\approx \mp^2 H_0^2.
\ee
Therefore, 
\be
M\sim 1.88\times 10^{4.75}\beta^{-1/8}(-\ln 10^{-16}\eta)^{1/2}~{\rm GeV},
\ee
i.e. $M\sim 10^5$ GeV (which is insensitive 
in order of magnitude to the parameters 
$10^{-5}\lsim\eta\sim 1.0$ and $0.5\lsim\beta\lsim 1.0$). 
This result is similar to that of \cite{peebles2}. We have 
used $\nu\approx 0.64$ and the present time Hubble constant 
$H_0\approx 1.68\times 10^{-42}$ GeV.   

Also, for this order of magnitude of $M$ we obtain
\be
\frac{a_\phi}{a_*}\sim 10^{19}.
\ee
Therefore, the condition $a_\phi>a_r$ results in $\eta\gsim 10^{-5}$. 
In regions where the particle creation is below this threshold there 
is not enough matter to sustain a radiation-dominated phase, and the 
dilaton moves on from a KE-dominated era straight into 
a DE-dominated one. 

\subsection{Bounds on $\xi$}

The value of the dilaton in the $D_-$ regions at present is 
\be 
\phi_r\approx (5.0-6.0)\times 10^{19}~{\rm GeV},
\ee
and from (\ref{planckmass}) we see this can pose a problem 
for the order of magnitude of $\xi$. We have 
$\phi_r/\mp\approx 15$, so in order to 
avoid having a catastrophic $m_{\rm p,eff}\sim 0$, we 
must have $\xi\lsim 1.7\times 10^{-3}$, which is one order of magnitude 
below the observational estimates we have cited. 
No such bound exists for $\xi<0$. 

\begin{figure}[t]
\centering
\leavevmode\epsfysize=6.3cm \epsfbox{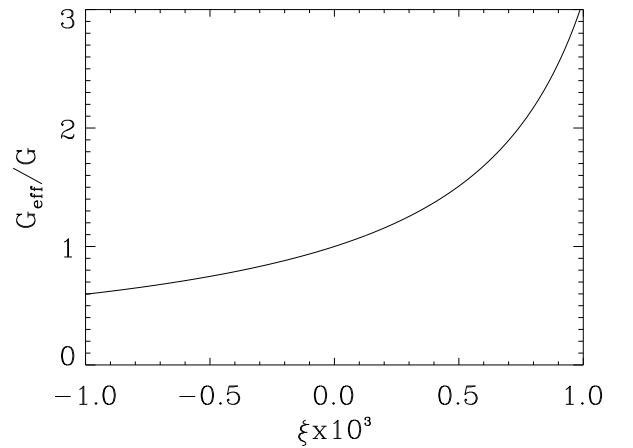}\\
\vskip 0.2cm
\caption[fig5]{Ratio of $G_{\rm eff}$ in accelerating regions with 
respect to $G$ measured in the trough. $\xi>0$ opens up the 
possibility of a value of $G_{\rm eff}$ several times greater than 
that measured locally. For $\xi<0$ this ratio is about half.}
\end{figure}

One unusual feature of $D_-$ regions is that $m_{\rm p,eff}$ 
(and therefore $G_{\rm eff}$) differs with respect 
to previous cosmic eras, where $m_{\rm p,eff}\approx \mp$ 
throughout. The value of $m_{\rm p,eff}$ 
in $D_-$ regions today differs 
from the value measured in galaxies ($D_+$ regions).  
Therefore, from the value of the Planck mass measured 
locally we can derive $m_{\rm p,eff}$ in regions that are
presently accelerating. For $\xi= 1.0\times 10^{-3}$ 
($-1.0\times 10^{-3}$) this results in  
$m_{\rm p,eff}\approx 2.39\times 10^{18}$ GeV ($5.44\times 10^{18}$
GeV), or equivalently $G_{\rm eff}/G \approx 3.07 (0.60)$.

\section{Dilaton mass}

The mass of the oscillating dilaton is 
\be
m_\phi\equiv \sqrt{|V^{\prime\prime}(\phi_0)|},
\ee
which contributes, together with the $\rhom$ of the particles 
produced after EOI, to the DM content. In our case, 
\be
m_\phi \approx 
\frac{2\sqrt{2}\lambda^{1/2}M^6}{\phi_0(\phi_0^8+M^8)^{1/2}},
\ee
or equivalently
\be \label{pmass}
m_\phi\approx 2\lambda^{1/2}\nu^{-1}M.
\ee
Therefore, considering $0<\nu<1$, 
the lightest dilaton  
that we can have is $m_\phi\approx 20~{\rm MeV}$ (hence a 
WDM-like behaviour with $m_\phi\sim 1~{\rm keV}$ is out of 
the question), and heavier masses are obtained for $\nu <1$, as 
shown in Fig.6. The limit $\nu\to0$ 
is catastrophic in that $V_{\rm hill}\to\infty$ and
$m_\phi\to\infty$. 

Cho \& Keum \cite{cho} have investigated the bounds on the 
dilaton mass if it is to fulfill the role of dominant source 
of DM in the universe. They conclude that there are two relevant mass ranges, 
$m_{\phi_1}\approx 0.5~{\rm keV}$ 
and $m_{\phi_2}\approx 270~{\rm MeV}$. The underlying argument is 
that if $m>m_{\phi_2}$, then the dilaton does not survive long 
enough to dominate DM, and on the other hand if $m<m_{\phi_1}$, 
it does survive long enough but it is not heavy enough to dominate 
DM. They also find that the dilaton of mass $m_{\phi_2}$ has 
a free streaming distance $\sim 7.4~{\rm pc}$ and is as a consequence 
a good candidate for CDM. 

In our model, if we wish the dilaton in the trough to behave like 
CDM within the parameters of \cite{cho}, this entails 
$\nu\approx 7.4\times 10^{-2}$. 
In this case the height of the hill is  
$V_{\rm hill}\approx 3.0\times 10^{10}~{\rm GeV}^4$, and relative to the 
energy density of the dilaton 
\be
\frac{V_{\rm hill}}{\rhop}\sim 10^{-51},
\ee
so the barrier of the potential is extremely low for the 
KE of the field, and hence the model can only succeed if there 
is abundant KE loss for the dilaton to be trapped in the trough. 

If we adopt Cho \& Keum's $m_{\phi_2}$ as an upper limit for the dilaton, then 
the allowed range in our model is 
$-1.70\lsim {\rm log}(m_\phi/{\rm GeV})\lsim -0.57$. One can 
consider a dilaton at the lower end of this interval, 
for instance $\nu\approx 0.64$ 
(just below unity to have enough of a trough, as discussed in 
\S II), i.e. 
$m_\phi\approx 31~{\rm MeV}$, with a free streaming distance 
$\sim 65~{\rm pc}$. This facilitates erasing small-scale structure 
by an order of magnitude over and above CDM. 

\begin{figure}[t]
\centering
\leavevmode\epsfysize=6.3cm \epsfbox{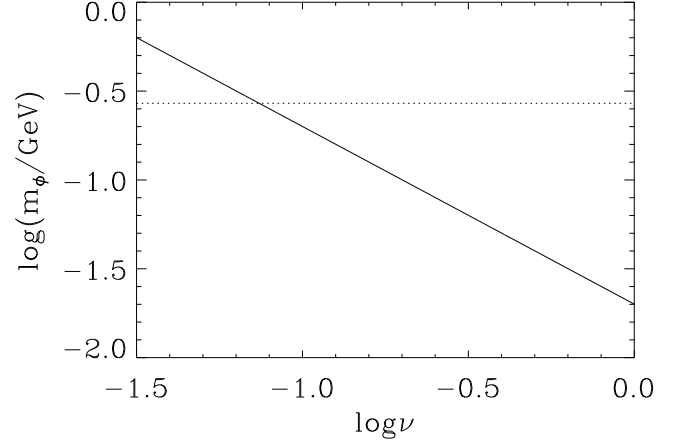}\\
\vskip 0.2cm
\caption[fig6]{Dilaton mass in $D_+$ regions or ``solitonic halos'' 
given by (\ref{pmass}) ({\it solid line}). Upper bound ({\it dotted 
line}) corresponds to Cho \& Keum's $m_{\phi_2}=270~{\rm MeV}$.}
\end{figure}

\section{Gravitational cooling} 

As first observed by Seidel \& Seun \cite{seidel2}, 
a scalar field perturbation can dissipate KE efficiently 
by ``gravitational cooling''. Seidel \& Seun tackled the 
problem of scalar field perturbations collapsing into 
boson stars to explain DM. Without 
a cooling mechanism, the perturbation remains a very diffuse 
virialised cloud. They proposed that in a process similar 
to the violent relaxation of stellar systems (see e.g. 
Spergel \& Hernquist \cite{spergel2}), the perturbation  
ejects part of the scalar field, carrying the excess KE, and then 
forms a star by means of its own self-gravity. 

In our scenario the dilaton rolls down the 
trough of the potential soon after EOI, and due to the steepness 
of the potential, it gains too much KE to remain 
there. There is some KE loss 
due to reheating but this is not enough to confine it 
to the trough. Gravitational cooling removes the KE 
excess from random regions passing it on to adjacent ones. It 
is not unreasonable to expect that the inhomogeneities 
caused by $\eta$ during particle production enhance the 
gravitational collapse of the dilaton and therefore 
gravitational cooling in regions were $\eta$ is greater is 
favoured. These 
(which we have called $D_+$ regions) then become ``oscillatons'', 
which are in essence extended halos where galaxies may form. 
This type of solution has been studied 
by Seidel \& Suen \cite{seidel} and also Ure\~{n}a-L\'{o}pez 
\cite{urena2}), adopting spherical symmetry, 
\be
\phi(t,r)=\sum_{n=0}^{\infty}\phi_n(r) \cos
\big[ (2n+1)\omega_0 t\big],
\ee
where $\phi_n$ can be computed numerically given boundary conditions 
of continuity and non-singular behaviour. 

The regions adjacent to the oscillaton carry the excess of KE 
and soon move out of the trough. Owing to the large transfer of KE, these 
$D_{-}$ regions have a larger $\rhop$ than that given by 
(\ref{KE}), under the simplified assumption of 
the homogeneous background evolution.  

\subsection{Domain formation}

For the parameters we have considered, $\phi_0\approx 10^5~{\rm GeV}$ 
and $\phi_r\approx 10^{19}~{\rm GeV}$, so there is a difference 
of fourteen orders of magnitude in $\phi$ between $D_-$ and 
$D_+$ regions. These regions are separated by domain walls where 
the field is near-stationary at various intermediate locations. 
In the $D_+$ domains $\pp\approx 0$ and 
\be 
\rhop\approx 2V(\phi)\approx \lambda\nu^{-4}(\phi^2-\phi_0^2)^2,  
\ee
is the (periodic) energy density of the oscillaton. 
In the $D_-$ domains, $\rhop$ and $\pp$ are dominated by 
the potential and KE is only residual, as described in the next 
section.

\begin{figure}[t]
\centering
\leavevmode\epsfysize=6.3cm \epsfbox{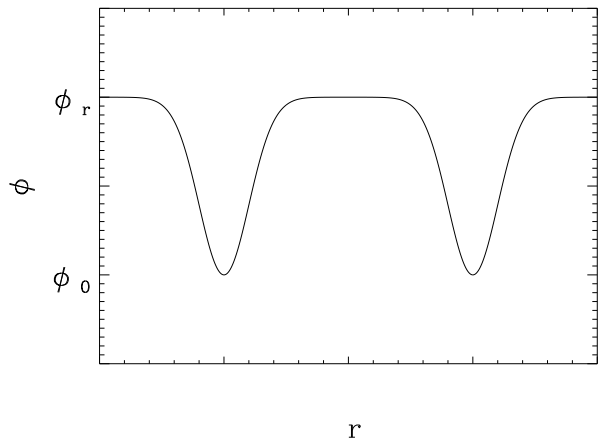}\\
\vskip 0.2cm
\caption[fig7]{$D_+$ (bottom) and $D_-$ (top) 
domains in one spatial dimension. $D_-$ domains undergo accelerated 
expansion whereas $D_+$ domains follow a powerlaw 
expansion.}
\end{figure}

Fig.6 shows qualitatively two $D_+$ domains immersed in a 
DE-dominated background. The dilaton at the minima oscillates 
around $\phi_0$, behaving as a dust-like non-relativistic 
matter fluctuation and the (local) expansion factor in these domains is 
$a\sim t^{2/3}$. By contrast, $D_-$ regions undergo accelerated  
expansion as in a DE-driven universe. This suggests that most 
physical space is occupied by $D_-$ regions, however few of these 
there are, 
even if DE-domination started in the very recent past. The field 
equation (\ref{KG}) contains strong spatial gradients 
for the domain walls and certain assumptions need to be made 
about the symmetries (form of the metric) and the DM 
distibution in the domain in order to compute solutions. 
One such assumption can be to impose upon the solutions 
the condition to emulate flat rotation curves 
of galaxies, following for instance the method of Matos, 
Guzm\'{a}n \& N\'{u}\~{n}ez \cite{matos2}. We hope to report on 
a study along these lines in the near future.

\section{Jeans length} 

The dilaton in the $D_+$ regions is homogeneous 
with small perturbations in a time-dependent background. Following 
the work of Khlopov, Malomed \& Zel'dovich \cite{khlopov}, 
the gravitational instability of such a configuration 
(the oscillaton) is similar to the Jeans instability of 
dust-like non-relativistic matter. 

The field equation (\ref{KG}) becomes 
\be \label{fieldD}
\Box\phi +6\xi (\dot H+2H^2)\phi 
+\lambda\nu^{-4}(\phi^2-\phi_0^2)\phi =0.
\ee
In the Newtonian approximation, 
\be \label{newtonian}
\ddot\phi-\Delta\phi-m_\phi^2(1+2\varphi)\phi
+\lambda\nu^{-4} \phi^3=0,
\ee
where the gravitational potential $\varphi$ is caused 
by the perturbations and it satisfies Poisson's equation 
\be \label{poisson}
\Delta\varphi=4\pi G \Big[ T_{00}(\phi)-T_{00}(\langle\phi\rangle)\Big].
\ee
Assuming a background solution 
$\langle\phi\rangle =\phi_0\cos \omega t$, we have from (\ref{newtonian}) 
\be 
\omega^2=5\lambda \nu^{-2} M^2.
\ee
We consider the perturbed solution 
\be \label{pert1}
\phi = \tilde\phi(t,x_i) \cos [\omega t +\tau(t,x_i)],
\ee
where 
\be 
\tilde\phi=\phi_0+\phi_1\exp (\Omega t+i k_jx^j),
\ee
\be 
\tau=\tau_1\exp (\Omega t+i k_jx^j),
\ee
\be \label{pert2}
\varphi=\varphi_1 \exp(\Omega t+i k_jx^j). 
\ee
Substituting (\ref{pert1})-(\ref{pert2}) in 
(\ref{newtonian})(\ref{poisson}), we obtain 
after a bit of algebra the dispersion relation
\[ 
(\Omega^2+k^2+2\lambda\nu^{-2}M^2)(\Omega^2+k^2)
\]
\be
+4\omega^2\Omega^2
-320\pi G \lambda^2\nu^{-2} M^6=0,
\ee
and therefore the Jeans wave-number becomes 
\be
k_J^2\sim 160\pi\lambda M^4.
\ee

In Fig.8 we show the Jeans length $\lambda_J$ for the 
range of dilaton masses considered in \S VI, i.e. 
$20~{\rm MeV}\lsim m_\phi\lsim 270~{\rm MeV}$. The Jeans length 
for these masses falls within the range  
$0.24~{\rm cm}\lsim \lambda_J\lsim 268~{\rm cm}$. We plot $\lambda_J$ 
along with the Compton length $\lambda_c\sim m_\phi^{-1}$, and 
it is apparent the latter is smaller by 12-14 orders of magnitude. 
Clearly, such a small $\lambda_J$ leaves the dilaton in no difficulty 
to form stable solitonic halos that constitute the bulk of DM in 
galaxies.

\begin{figure}[t]
\centering
\leavevmode\epsfysize=6.3cm \epsfbox{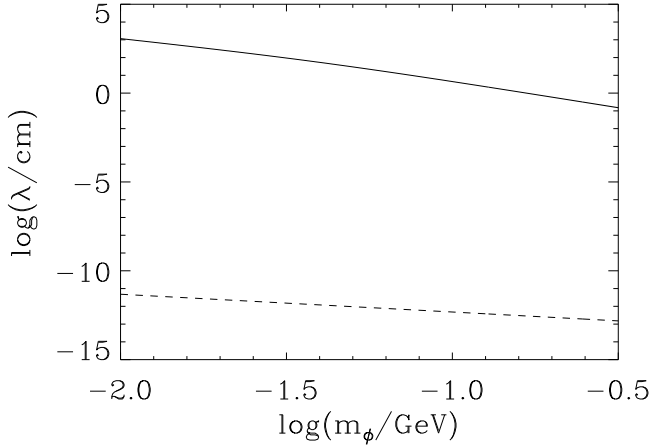}\\
\vskip 0.2cm
\caption[fig8]{Jeans length ({\it solid line}) and 
Compton length ({\it dashed line}).}
\end{figure}

\section{Residual KE}

So far we have considered that $\opp$ evolves from 
$1$ in the KE-dominated era, to $-1$ 
during DE domination. The transition entails an ever-diminishing KE, until 
it becomes negligible at $\phi\approx \phi_r$. 
However, observationally 
$\langle \opp\rangle = -0.65\pm 0.07$ (e.g. by recent CMB data 
\cite{WMAP,DASI}), and in order to reproduce 
this a residual KE must exist at present. 

We consider the large-scale averaged quantities 
\be 
\langle\rhop\rangle\approx 
\sigma\Big\langle\frac{1}{2}\dot\phi^2\Big\rangle +V(\phi_r),
\ee 
\be
\langle\pp\rangle\approx 
\sigma\Big\langle\frac{1}{2}\dot\phi^2\Big\rangle -V(\phi_r),
\ee
where $\sigma$ is a large number (reflecting the KE transfer 
from $D_+$ regions to $D_-$), and 
$V(\phi_r)\gg \sigma\langle\dot\phi^2/2\rangle$. 
A further contribution to the 
residual KE comes from the $\xi$ terms on the RHS of
(\ref{rho})(\ref{pressure}). Therefore
\be
\sigma\approx \frac{V(\phi_r)}{\langle\dot\phi^2/2\rangle}
\frac{(1+\langle\opp\rangle)}{(1-\langle\opp\rangle)},
\ee
and from (\ref{KE}) 
\be 
\Big\langle\frac{\dot\phi^2}{2}\Big\rangle _{\rm today}
\sim \lambda \mp^4\,10^{-96}.
\ee
For the sake of the example we take $\langle\opp\rangle= -0.65$, hence 
\be
\sigma\sim 10^{25}.
\ee
This is a very large factor that lends support to the 
hypothesis that in fact most regions of the universe are 
$D_+$ and their KE is passed on to very few $D_-$ regions. 
If we consider that in the neighbourhood of the trough 
$\rhom/\rhop\ll 1$, and $D_+$ regions transfer all KE 
to $D_-$, then (within the coarse approximation of 
the background evolution equations) 
we have one $D_-$ per $10^{25}$ oscillatons.

\section{Conclusions}

We have studied a cosmological model of an inhomogenous 
dilaton where 
DM and DE originate from the same source. The simplicity of 
the model is attractive mainly in two respects: firstly,   
the dilaton, a physically-motivated scalar field, triggers 
the inflationary process without the assistance of additional 
scalar fields; secondly, it is not surprising that the energy 
density of DM and DE are of the same order of magnitude as they 
are caused by the same field. There is good reason then to 
believe that there are fewer grounds to assert there is 
a so-called ``coincidence problem''.  

The model predicts that there is little or no  
DM outside the $D_+$ regions, other than $\rhom$ produced 
at the beginning of the KE-dominated era. If individual galaxies 
are to be identified with $D_+$ regions, and the space between 
them is $D_-$, then one 
does not expect to find large amounts of DM in the intergalactic 
medium. Having said that, there is no reason to expect that 
each and every soliton ultimately produces a galaxy, and 
therefore dark $D_+$ regions may also be present in the 
intergalactic medium, which are technically detectable by 
lensing experiments.   

To construct the model we have used a potential with 
a ``trough'' feature, inspired by previous models that  
result in a DE-dominated universe, in particular the ``quintessential 
inflation'' model by Peebles \& Vilenkin \cite{peebles2}. 
The trough feature is crucial to give way to solitonic solutions, 
as is the process of gravitational cooling, proposed by 
Seidel \& Suen \cite{seidel2}. Once again, like with most 
cosmological models, the reliance on the features of the 
potential to produce the desirable solutions makes 
it the ``weakest link'', in that its acceptance is inescapably  
ad hoc. It would be more desirable to derive $V$ from fundamental 
arguments. 

The inhomogeneous dilaton opens up the possibility of large variations 
of $G$ in $D_-$ regions. Accurate predictions are $\xi$-dependent, 
though in this 
paper we have looked into the possibility that whereas $G$ has 
changed little within $D_+$ regions, it can be several times 
greater(smaller) in the intergalactic medium ($D_-$ regions) 
if $\xi$ is positive(negative). This is provided we accept the identification 
of $D_+$ regions with galaxies, which we must stress is not an imperative of 
the model, though an interesting possibility.


\begin{thebibliography}{999}

\bibitem{peebles2}
P.J.E. Peebles and A. Vilenkin, \prd {\bf 59}, 063505 (1999)

\bibitem{ford}
L.H. Ford, \prd {\bf 35}, 2955-2960 (1987)

\bibitem{spokoiny}
B. Spokoiny, Phys. Lett. B {\bf 315}, 40-45 (1993)

\bibitem{seidel2}
E. Seidel and W-M. Suen, Phys. Rev. Lett. {\bf 72}, 2516 (1994)

\bibitem{perrotta3}
F. Perrotta, C. Baccigalpi and S. Matarrese, \prd {\bf 61}, 023507
(1999)

\bibitem{amendola2}
L. Amendola, \prd {\bf 62}, 043511 (2000)

\bibitem{baccigalupi2}
C. Baccigalupi, S. Matarrese and F. Perrotta, \prd {\bf 62}, 
123510 (2000)

\bibitem{bartolo}
N. Bartolo and M. Pietroni, \prd {\bf 61}, 023518 (1999)

\bibitem{chiba}
T. Chiba, \prd {\bf 60}, 083508 (1999)

\bibitem{chiba2}
T. Chiba, \prd {\bf 64}, 103503 (2001)

\bibitem{esposito}
G. Esposito-Far\`{e}se and D. Polarski, \prd {\bf 63}, 063504 (2001)

\bibitem{perrotta2}
F. Perrotta and C. Baccigalupi, \prd {\bf 65}, 123505 (2002) 

\bibitem{riazuelo}
A. Riazuelo and J.P. Unzan, \prd {\bf 62}, 083506 (2000)

\bibitem{ritis}
R. de Ritis, A.A. Marino, C. Rubano and P. Scudellaro, \prd {\bf 62}, 
043506 (2000)

\bibitem{uzan}
J.P. Uzan, \prd {\bf 59}, 123510 (1999)

\bibitem{padmanabhan}
T. Padmanabhan, \prd {\bf 66}, 021301 (2002)

\bibitem{bagla}
J.S. Bagla, H.K. Jassal and T. Padmanabhan, \prd {\bf 67}, 
063504 (2003) 

\bibitem{padmanabhan2}
T. Padmanabhan and T.R. Choudhury, \prd {\bf 66}, 081301 (2002) 

\bibitem{birrell}
N.D. Birrell and P.C.W. Davis, {\it Quantum Fields in Curved Spacetime}, 
Cambridge Monographs on Mathematical Physics, 
Cambridge University Press, Cambridge (1982)

\bibitem{maeda}
K. Maeda, Class. Quantum Grav. {\bf 3}, 233 (1986)

\bibitem{WMAP}
C.L. Bennett et al., arXiv:astro-ph/0302207 (Preliminary maps and 
basic results); H.V. Peiris et al., arXiv:astro-ph/0302225
(Implications for inflation); D.N. Spergel et al., 
arXiv:astro-ph/0302209 (Determination of cosmological parameters);
WMAP technical 
and complete bibliography 
http://lambda.gsfc.nasa.gov/product/map/
map$\_$bibliography.html

\bibitem{DASI}
J.M. Kovac, E.M. Leitch, C. Pryke, J.E. Carlstrom, N.W. Halverson 
and W.L. Holzapfel, Nature (London) {\bf 420}, 772-287 (2002); 
C. Pryke, N.W. Halverson, E.M. Leitch, J. Kovac, J.E. Carlstrom, 
W.L. Holzapfel and M. Dragovan, Astrophys. J. {\bf 568}, 46-51 (2002)

\bibitem{lee}
A.T. Lee et al., Astrophys. J. Lett. {\bf 561}, L1 (2001)

\bibitem{stompor}
R. Stompor et al., Astrophys. J. Lett. {\bf 561}, L7 (2001)

\bibitem{bernardis}
P. de Bernardis et al., Astrophys. J. {\bf 564}, 559 (2002)

\bibitem{hu3}
W. Hu and S. Dodelson, S. Annu. Rev. Astron. Astrophys. {\bf 40}, 
171-216 (2002)


\bibitem{linde}
A.D. Linde, ``Particle Physics and Inflationary Cosmology'' 
(Harwood, Chur, 1990)

\bibitem{carroll}
S.M. Carroll, Phys. Rev. Lett. {\bf 81}, 3067-3070 (1998)

\bibitem{coble}
K. Coble, S. Dodelson and J.A. Frieman, \prd {\bf 55}, 1851-1859
(1997)

\bibitem{steinhardt}
P.J. Steinhardt, L. Wang and I. Zlatev, \prd {\bf 59}, 123504 (1999)

\bibitem{ratra}
B. Ratra and P.J.E. Peebles, \prd {\bf 37}, 3406-3427 (1988)

\bibitem{wetterich}
C. Wetterich, Nucl. Phys. {\bf B302}, 645 (1988)

\bibitem{zlatev}
I. Zlatev, L. Wang and P.J. Steinhardt, Phys. Rev. Lett.~{\bf 82},
896-899 (1999)

\bibitem{abreu}
J.P. Abreu, P. Crawford and J.P. Mimoso, Class. Quantum Grav. 
{\bf 11}, 1919-1939 (1994)

\bibitem{accetta}
F.S. Accetta and J.J. Trester, \prd {\bf 39}, 2854 (1989) 

\bibitem{amendola}
L. Amendola, \prd {\bf 60}, 043501 (1999)


\bibitem{damour2}
T. Damour and G. Esposito-Far\`{e}se, Phys. Rev. Lett. {\bf 70}, 2220-2223 
(1993); \prd {\bf 54}, 1474-1491 (1996); ibid. {\bf 58}, 042001 (1998) 

\bibitem{hwang}
J.C. Hwang, Class. Quantum Grav. {\bf 7}, 1613-1631 (1990); 
Astrophys. J. {\bf 375}, 443 (1991); \prd {\bf 53}, 762 (1996) 

\bibitem{nagata}
R. Nagata, T. Chiba and N. Sugiyama, \prd {\bf 66}, 103510 (2002)

\bibitem{gillies}
G.T. Gillies, Rep. Prog. Phys. {\bf 60}, 151 (1997)

\bibitem{joyce}
M. Joyce, \prd {\bf 55}, 1875 (1997)

\bibitem{cho}
Y.M. Cho, Y.Y. Keum, Mod. Phys. Lett. A {\bf 13}, 109 (1998)

\bibitem{spergel2} 
D.N. Spergel and L. Hernquist, Ap. J. {\bf 397}, L75 (1992) 

\bibitem{seidel}
E. Seidel and W-M. Suen, Phys. Rev. Lett. {\bf 66}, 1659 (1991)

\bibitem{urena2}
L.A. Ure\~{n}a-L\'{o}pez, Class. Quantum Grav. {\bf 19}, 2617-2632
(2002) 

\bibitem{damour4}
T. Damour, A. Vilenkin, \prd {\bf 53}, 2981 (1996)

\bibitem{giovannini}
M. Giovannini, \prd {\bf 58}, 083504 (1998)

\bibitem{matos2}
T.Matos, F.S. Guzm\'{a}n and D. N\'{u}\~{n}ez, \prd {\bf 62}, 
061301 (2000)

\bibitem{khlopov}
M.Yu. Khlopov, B.A. Malomed and Ya.B. Zel'dovich, MNRAS {\bf 215}, 
575 (1985)

\end{thebibliography}
\end{document}